\let\ifarxiv=\iftrue     % ARXIV VERSION
\pdfoutput=1

\ifarxiv

\documentclass[12pt,a4paper]{article}
\usepackage[a4paper,text={450pt,650pt},centering]{geometry}

\fi

\ifarxiv\else

\documentclass[11pt,a4paper]{article}
\usepackage{mathptmx}
\usepackage[a4paper,text={130mm,198mm}]{geometry}

\fi

\usepackage{amsmath,amssymb}
\usepackage[bookmarks=true,hyperfigures=true]{hyperref}
\usepackage{graphicx}
\usepackage[nosort]{cite}
\usepackage{color}
 \usepackage{bbm}

\newcommand{\beq}{\begin{equation}}
\newcommand{\eeq}{\end{equation}}
\newcommand\beqa{\begin{eqnarray}}
\newcommand\eeqa{\end{eqnarray}}
\newcommand\bea{\begin{array}}
\newcommand\eea{\end{array}}

\newcommand{\nn}{\nonumber}
\newcommand{\la}[1]{\label{#1}}

\newcommand{\Tr}{{\rm Tr}}

\def\[{\left[}
\def\]{\right]}

\def\s{\sigma}

\def\({\left(}
\def\){\right)}
\def\[{\left[}
\def\]{\right]}

\def\<{\langle}
\def\>{\rangle}

    \def\be{\begin{eqnarray}}
    \def\ee{\end{eqnarray}}
    \def\no{\nonumber}
    \def\bn{\begin{enumerate}}
    \def\en{\end{enumerate}}
    \def\({\left(}
    \def\){\right)}
    \def\<{\left\langle\,}
    \def\>{\, \right\rangle}
    \def\[{\left[}
    \def\]{\right]}
    \def\la{\label}
    \def\g{\gamma}
    \def\CO{{\cal O}}
    \def\pd{\partial}
    \def\MZ{{\mathbb{Z}}}
    \def\sign{{\rm{sign}}\,}
    \def\xpm{ x^{\pm} }
    \def\xp{ x^{+} }
    \def\xm{ x^{-} }
    
    \def\yp{ y^{+}}
    \def\ym{ y^{-}}

\let\oldbfseries=\bfseries
\let\oldmdseries=\mdseries
\let\oldnormalfont=\normalfont
\renewcommand{\bfseries}{\oldbfseries\boldmath}
\renewcommand{\mdseries}{\oldmdseries\unboldmath}
\renewcommand{\normalfont}{\oldnormalfont\unboldmath}

\allowdisplaybreaks[3]

\numberwithin{equation}{section}

\usepackage[font=small,labelfont=bf,width=0.85\textwidth]{caption}

\providecommand{\hypersetup}[1]{}
\providecommand{\texorpdfstring}[2]{#1}

\hypersetup{plainpages=false}
\hypersetup{pdfpagemode=UseNone}
\hypersetup{bookmarksnumbered=true}
\hypersetup{pdfstartview=FitH}
\hypersetup{colorlinks=false}
\hypersetup{citebordercolor={.5 1 .5}}
\hypersetup{urlbordercolor={.5 1 1}}
\hypersetup{linkbordercolor={1 .7 .7}}
\providecommand{\arxivref}[2]{\href{http://arxiv.org/abs/#1}{#2}}

\providecommand{\href}[2]{#2}
\providecommand{\arxivlink}[1]{\href{http://arxiv.org/abs/#1}{arxiv:#1}}

\begin{document}

%%%%%%%%%%%%%%%%%%%%%%%%%%%%%%%%%%%%%%%%%%%%%%%%%%%%%%%%%%%%%%%%%%%%%%%%%%%%%%%%
%%%%%%%%%%%%%%%%%%%%%%%%%%%%%%%%%%%%%%%%%%%%%%%%%%%%%%%%%%%%%%%%%%%%%%%%%%%%%%%%
% TITLE PAGE

\thispagestyle{empty}
\phantomsection
\addcontentsline{toc}{section}{Title}

\begin{flushright}\footnotesize%
\texttt{\arxivlink{1012.3992}}\\
overview article: \texttt{\arxivlink{1012.3982}}%
\vspace{1em}%
\end{flushright}

\begingroup\parindent0pt
\begingroup\bfseries\ifarxiv\Large\else\LARGE\fi
\hypersetup{pdftitle={Review of AdS/CFT Integrability, Chapter III.3: The dressing factor}}%
Review of AdS/CFT Integrability, Chapter III.3:\\The dressing factor
\par\endgroup
\vspace{1.5em}
\begingroup\ifarxiv\scshape\else\large\fi%
\hypersetup{pdfauthor={Pedro Vieira, Dmytro Volin}}%
Pedro Vieira$^{a}$, Dmytro Volin$^{b}$ 
\par\endgroup
\vspace{1em}
\begingroup\itshape
\textit{$^a$
Perimeter Institute for Theoretical Physics, \\Waterloo,
Ontario N2L 2Y5 , Canada} \\
\textit{$^{b}$ Department of Physics, The Pennsylvania State University, \\University Park, PA 16802, USA} \\
\par\endgroup
\vspace{1em}
\begingroup\ttfamily
pedrogvieira AT gmail.com \qquad, \qquad  dvolin AT psu.edu
\par\endgroup
\vspace{1.0em}
\endgroup

\begin{center}
\includegraphics[width=5cm]{TitleIII3.mps}%figure for your chapter
\vspace{1.0em}
\end{center}

\paragraph{Abstract:}
We review the construction of the AdS/CFT dressing factor, its analytic properties and several checks of its validity.

\ifarxiv\else
%\paragraph{Mathematics Subject Classification (2010):} 
%...
% http://www.ams.org/msc
\fi

\ifarxiv\else
%\paragraph{Keywords:} 
%...
\fi

\newpage

%%%%%%%%%%%%%%%%%%%%%%%%%%%%%%%%%%%%%%%%%%%%%%%%%%%%%%%%%%%%%%%%%%%%%%%%%%%%%%%%
%%%%%%%%%%%%%%%%%%%%%%%%%%%%%%%%%%%%%%%%%%%%%%%%%%%%%%%%%%%%%%%%%%%%%%%%%%%%%%%%
% BODY

%\tableofcontents
\section{Introduction} \la{Intro}
The two-body S-matrix is the main figure in the solution of 1+1 dimensional integrable theories. \textit{In principle}, given this object there is a more or less well defined route to compute the asymptotic and even the finite-size spectrum of the theory.

Typically, particles have polarizations and therefore the S-matrix has a nontrivial matrix structure.
The various ratios of the S-matrix elements give us the relative weights of different scattering processes in the same theory. These ratios are mostly kinematic and highly constrained by the symmetries of the system. For integrable models, symmetry together with the condition of factorized scattering is usually enough to constraint completely the matrix structure of the S-matrix up to an overall scalar factor called \textit{dressing factor}. This factor contains much more dynamical information and is therefore considerably harder to derive. Usually, to find the correct dressing factor, one needs to consider unitarity and crossing equations supplemented by the knowledge of the exact bound state spectrum.

In this review we will study  the dressing factor for a very interesting integrable model which appears in the
spectrum problem of planar AdS/CFT \cite{chapIntro,chapSMat}.
%For short we will often denote this model by the \textit{AdS/CFT system}.

We start by considering (section \ref{O4sec}) a warm-up toy model, the $O(4)$ sigma model, which will be quite instructive. Then we move to the AdS/CFT system (sections \ref{crosssec} and \ref{expsec}). Our logical flow will be roughly the opposite of the chronological one. We will start by solving directly the crossing relation in section \ref{crosssec}. Then, in section \ref{expsec}, we will consider several rewritings and expansions of the dressing factor, including the Beisert-Eden-Staudacher solution \cite{Beisert:2006ez,Beisert:2006ib} and several weak and strong coupling expansions. Along the way we will describe analytical properties and remarkable checks of the AdS/CFT dressing factor which  established its correctness beyond reasonable doubt. In section \ref{conclusions} we present some concluding remarks.

\section{Dressing factor in the \texorpdfstring{$O(4)$}{O(4)} sigma model} 
\begin{figure}[t]
\centering
\includegraphics[width=10.00cm]{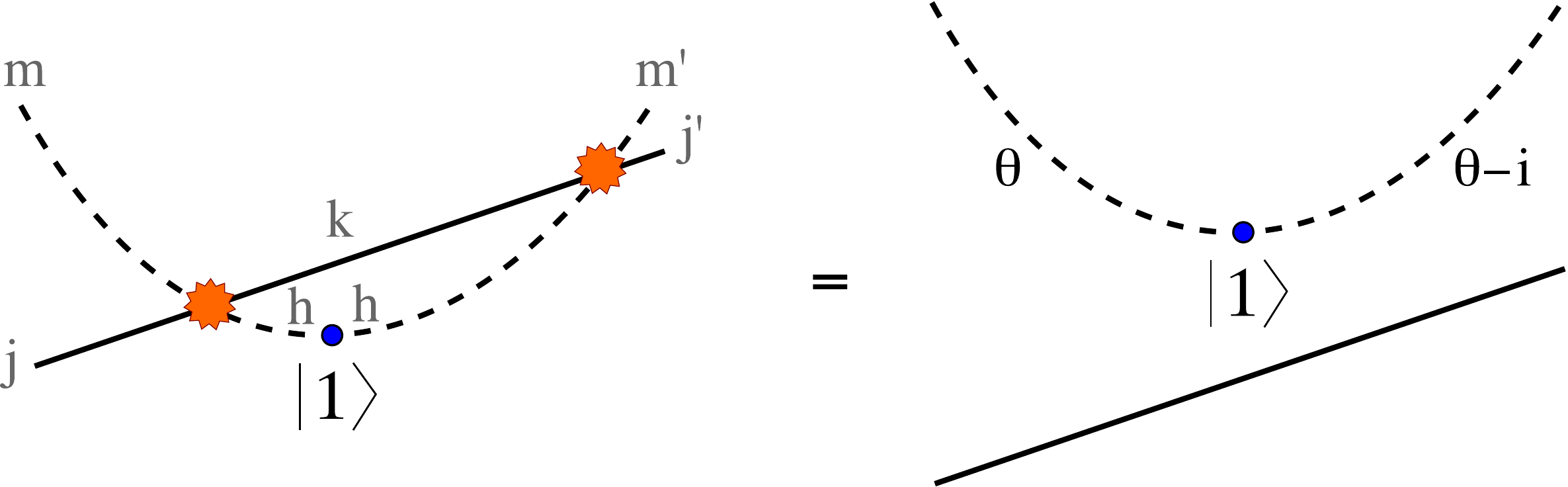}
\caption{ Scattering of a physical particle (solid line) with a composite of particle/antiparticle of zero total zero charges (dashed line) should be inconsequential \cite{Beisert:2005tm}. This picture translates into formula (\ref{crossingM}). }\label{crossingfig}
\end{figure}

\la{O4sec}
The $O(4)$ sigma model in two dimensions is an integrable relativistic model where particles have energy and momentum given by
\beq
\epsilon(\theta)=m \cosh(\pi\theta) \,\, , \,\, p(\theta)=m\sinh(\pi\theta) \,, %\nn
\eeq
where $\theta$ is the so called rapidity. Lorentz boosts amount to simple translations in $\theta$. Hence, the $O(4)$ Lorentz invariant two-body S-matrix must be a function of the difference of rapidities $\theta=\theta_1-\theta_2$ of the two particles being scattered, $\hat S_{2\to 2} (\theta_1,\theta_2)= \hat S(\theta)$. The hat emphasizes that this object is a \textit{matrix} since the scattered particles have isotopic degrees of freedom. Since the model is integrable the matrix structure of the S-matrix can be fixed from the symmetry of the problem together with the Yang-Baxter triangular relation \cite{Zamolodchikov:1977nu}. One finds
\beq
\[\hat S(\theta)\]_{hk}^{jl}=\sigma^2(\theta )\(\frac{i\theta  }{ (\theta -i)^2}\,\delta _{hk} \delta _{jl}+\frac{\theta  }{\theta -i} \,\delta _{hj} \delta _{lk}-\frac{i }{\theta-i } \,\delta _{hl} \delta _{jk}\) \,. %\nn
\eeq
The overall \textit{dressing factor} $\sigma^2(\theta)$ is however left undetermined since it drops out of Yang-Baxter. This \textit{function} is constrained by imposing extra physical conditions: crossing symmetry and unitarity. Let us adapt a nice argument by Beisert \cite{Beisert:2005tm} to derive the implications of crossing symmetry. We construct a composite singlet state with one particle and one antiparticle,
\beq
|1\rangle \propto \sum_{h=1}^4 |\{h,\theta\},\{h,\theta-i\}\rangle \la{singlet} \,.
\eeq
The second particle has the same color as the first particle and the opposite energy and momenta since $\theta\to \theta-i$ flips these two quantities. The singlet state (\ref{singlet}) is therefore a spurious state which has zero color, momentum and energy.  Physically we think of it as a virtual pair of particle and anti-particle created created by a vacuum fluctuation. Scattering of a physical particle through this bound state should be inessential, see figure \ref{crossingfig}. Algebraically, the condition depicted in this figure translates into
\beq
\sum_{h=1}^4\sum_{k=1}^4  \[\hat S(\theta)\]_{j h}^{k m} \[\hat S(\theta-i)\]_{k h}^{j'm'} = \delta_{j}^{j'} \delta_{mm'}   \,.\la{crossingM}
\eeq
This condition then implies $\sigma^2(\theta)\sigma^2(\theta-i) \,\theta^2/(\theta+i)^2=1$ or
\beq
\sigma(\theta+i/2) \sigma(\theta-i/2) = \frac{\theta-i/2}{\theta+i/2} \,. \la{crossingO4}
\eeq
This is the crossing relation \cite{Zamolodchikov:1977nu}. We will now provide two derivations of the so-called \textit{minimal solution} to the equation (\ref{crossingO4}).

\subsubsection*{First derivation}
%{\bf First derivation.} 
We start by taking the logarithm and derivative of the crossing relation (\ref{crossingO4}):
\beq
K^++K^-=  \frac{1}{2\pi i} \(\frac{1}{\theta-i/2}-\frac{1}{\theta+i/2}\) \la{logcross} \,,
\eeq
where $K\equiv \frac{1}{2\pi i} \frac{d}{d\theta} \log \sigma(\theta)$ and $f^{\pm}=f(\theta\pm i/2)$.
Next we go to Fourier space, 
\beq
\mathcal{F}(K^{\pm} )(\omega) =\int\limits_{-\infty}^{\infty} d\theta\, e^{i\theta \omega} K(\theta\pm i/2) =e^{\pm \omega/2}  \int\limits_{-\infty\pm i/2}^{\infty\pm i/2} d\theta \,e^{i \theta \omega} K(\theta) \la{integral} \,.
\eeq
At this point we need some physical input about the particle content of the $O(4)$ sigma model \cite{Zamolodchikov:1977nu}. Since there are no bound states, $\sigma(\theta)$ should not have poles in the strip $-1/2<{\rm Im}(\theta)<1/2$. Unitarity $\hat S(\theta) \hat S(-\theta)=\mathbb{I}$ yields $\sigma^2(\theta)\sigma^2(-\theta)=1$ which implies the absence of zeros in the same strip. The absence of poles and zeros in this strip is often called the \textit{minimality condition}.

Assuming this condition to hold we can solve (\ref{crossingO4}) uniquely. Indeed, since $K(\theta)$ has no singularities in the strip $-1/2<{\rm Im}(\theta)<1/2$ we can deform the integral contour (\ref{integral}) back to the real axis and conclude that $\mathcal{F}(K^{\pm})(\omega)=e^{\pm \omega/2}\mathcal{F}(K)(\omega)$. The Fourier transform of (\ref{logcross}) then yields
\beq
\mathcal{F}(K)(\omega) = \frac{e^{-|\omega|/2} }{2 \cosh({\omega}/{2}) }\,. %\nn
\eeq
The Kernel $K(\theta)$ is now trivially computed as
\beqa
K(\theta) =\mathcal{F}^{-1} \[\, \frac{e^{-|\omega|/2} }{2\cosh(\omega/2)}\,\]= \frac{1}{2\pi i} \frac{d}{d\theta} \log\[\frac{1}{i}  \frac{\Gamma\(\,1-\frac{\theta}{2i}\,\)\Gamma\(\,\frac{1}{2}+\frac{\theta}{2i}\,\)}{\Gamma\(1+\frac{\theta}{2i}\,\)\Gamma\(\,\frac{1}{2}-\frac{\theta}{2i}\,\)} \] \la{resultO4} \,.
\eeqa
The quantity inside the logarithm can then be identified with the dressing factor $\sigma(\theta)$.
To fix the constant of integration we imposed the condition
\be\label{sigma0}
\sigma^2(0)=-1\,.
\ee
 This constraint simply states that there cannot be two identical particles in the theory, which is indeed the case for the $O(4)$ sigma model.
%\footnote{The factor of $i$ multiplying the ratio of $\Gamma$ matrices ensures that $\sigma^2(0)=1$ which is a property which we derive from the unitarity condition $\sigma(\theta)\sigma(-\theta)=1$ for $\theta=0$.}
%\subsubsection*{Second derivation}
%We convolute equation (\ref{logcross}) with the inverse shift operator $s\equiv {1}/({2\pi \cosh \theta })$ to get
%\beq
%s*\(\,K^++K^-\,\)=\int\limits_{-\infty}^{\infty}d\theta' \frac{K(\theta'+i/2)+K(\theta'-i/2)}{2\pi \cosh(\theta)} = \oint\limits_{\Gamma} \frac{d\theta}{2\pi i} \,\frac{K(\theta')}{\sinh(\theta-\theta')}
%\eeq
%where $\Gamma$ is the rectangle encircling the physical strip described above. Under the minimal choice assumption $K(\theta')$ has no singularities in this region and hence this integral can be computed by residues at $\theta'=\theta$ yielding simply $K(\theta)$. Hence
%$K =s*K_1
%$ which indeed yields the correct $O(4)$ dressing factor.
\subsubsection*{Second derivation}
Let us redo the above derivation avoiding passing to Fourier space. The argumentation now will be admittedly more involved
but it is also the most useful for the analogy with solving the AdS/CFT crossing equation. First, the crossing equation (\ref{crossingO4}) can be re-written as
\be
    \sigma^{D+D^{-1}}=\theta^{-(D-D^{-1})}  \la{crossing20} \,,
\ee
where $D=e^{\frac i2\pd_\theta}$ is the shift operator, $D f(\theta)\equiv f(\theta+i/2)$, and
$
f^{\mathcal{O}[D]}\equiv \exp\( \mathcal{O}[D] \log f \) \,.
$
Formally we might be tempted to solve (\ref{crossing20}) as
\be
\sigma(\theta)=\theta^{f[D]} \,\,\, \,\,\, \text{where}\,\,\,  \,\,\, f[D]=-\frac{D-D^{-1}}{D+D^{-1}} \,. \la{fo4}
\ee
However, we have to interpret this expression with care. E.g. if we naively expand $f[D]=1+2\sum_{n=1}^{\infty} (-1)^n D^{2n}$ we see that $\sigma^2(0)=0$. Similarly, if we expand this operator at \textit{large $D$} as $f[D]=-1-2\sum_{n=1}^{\infty} (-1)^n D^{-2n}$ we get $1/\sigma^2(0)=0$. In both cases we face an obvious contradiction with the minimality condition. The only possible interpretation of (\ref{fo4}) which is consistent with the minimality condition and (\ref{sigma0}) is
\be
f[D]=\frac{D^{-2}}{1+D^{-2}}-\frac{D^2}{1+D^{2}} \equiv \sum_{n=1}^\infty (-1)^{n} D^{2n} - \sum_{n=1}^\infty (-1)^{n}D^{-2n}\,,  \la{fo4good}
 \ee
so that
\beqa\label{sigma1}
    \sigma(\theta)&=&\exp\(\sum_{n=1}^\infty (-1)^{n} D^{2n}\log(\theta) - \sum_{n=1}^\infty (-1)^{n}D^{-2n}\log(\theta)  \)\\
&=&c_{reg}\prod _{n=0}^{\infty } \frac{(\theta + 2 ni ) (\theta -(2 n+1)i)}{(\theta -2 ni)(\theta +(2 n+1)i) }=   -c_{reg} \frac{\Gamma\(\,1-\frac{\theta}{2i}\,\)\Gamma\(\,\frac{1}{2}+\frac{\theta}{2i}\,\)}{\Gamma\(1+\frac{\theta}{2i}\,\)\Gamma\(\,\frac{1}{2}-\frac{\theta}{2i}\,\)} \,.\nn
\eeqa
The sums in the exponent in (\ref{sigma1}) are divergent and therefore need to be regularized. The simplest way to do regularization is by computing the derivative of these sums and then integrating back. This procedure introduces an unknown constant of integration: $c_{reg}$. Using (\ref{sigma0}) we fix it to $c_{reg}=i$. We recover precisely what we derived before in (\ref{resultO4}).

This derivation of the dressing factor is almost the same as the first one. Indeed, in both cases, to obtain regular expressions we need to take a derivative of the logarithm of the crossing equation. More importantly, the shift operator $D=e^{\frac i2\pd_\theta}$ is Fourier transformed to $\mathcal{F(D)}=e^{-\frac i2\omega}$. However, one advantage of using the second derivation is that (\ref{sigma1}) explicitly highlights the analytic structure of the solution, in particular by looking at its zeros and poles we immediately recognize the ratio of gamma functions. In the case of the AdS/CFT, when the analytical structure is more involved, use of shift operators instead of their Fourier transforms is even more preferable.

Let us comment on a trivial feature of the above derivations. The (logarithm of the) dressing factor is singular  for $\theta=\pm i$. This means that for practical purposes, it was a good idea to shift from the original functional equation $\sigma(\theta)\sigma(\theta-i)=\dots$ to the relation (\ref{crossingO4}) of the form $\sigma(\theta+i/2)\sigma(\theta-i/2)=\dots$. It avoids the burden of carrying around several $i0$'s in the above derivations.

\section{Crossing equation} 
\begin{figure}[t]
\centering
\includegraphics[width=11.00cm]{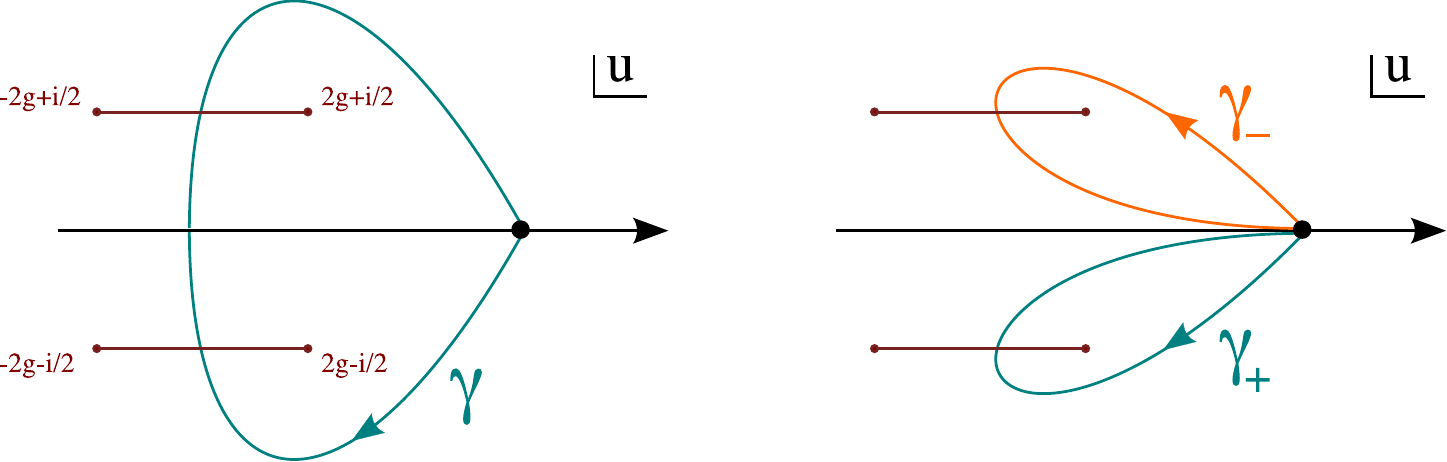}
\caption{Analytical continuation contours used in the crossing equation. The horizontal segments corresponds to the choice of branch-cuts for $x^{\pm}(u)=x(u\pm i/2)$ such that $|x(u)|>1$.  The upper (lower) cut is the cut of $x^{-}$ ($x^{+}$).}\label{fig:gammacross}
\end{figure}

\la{crosssec}
The AdS/CFT system is considerably more complicated than the $O(4)$ sigma model described above. However many features have a clear analogue in both models. The energy and momenta of the particle excitations, which are also called magnons, are now given by \cite{Santambrogio:2002sb,Beisert:2004hm,Beisert:2005tm}
\be
 e^{ip}=\frac{\xp}{\xm},\ \ \ \ \
 E=ig\left(\xm-\frac 1{\xm}-\xp+\frac 1{\xp}\right)\,, \la{dispersionAdS}
%\,\, E=1+2ig\left(\frac 1{\xp}-\frac 1{\xm}\right). \la{dispersionAdS}
\ee
where $\xpm\equiv x(u\pm i/2)$, $u/g\equiv x(u)+1/x(u)$ and $g\equiv \sqrt{\lambda}/4\pi$. The variable $u$ is the Bethe rapidity and is the direct analogue of $\theta$ in the $O(4)$ sigma model; $x(u)$ is the so called Zhukovsky variable. Unless otherwise stated, we choose the branch of the Zhukovsky variables such that $|\xpm(u)|>1$. This is the so called physical region, the one with a good $g\to 0 $ limit.
We represent the cuts of the functions $x^{\pm}(u)$ as uniting the corresponding branch-points horizontally in the complex $u$ plane, see figure \ref{fig:gammacross}.

The two-body  scattering matrix of the AdS/CFT  system, $\hat S=\hat S_0\times \s^{-2}$, depends on the rapidities of the particles being scattered and on the 't Hooft coupling. As in the previous example, we have a matrix part $\hat S_0$ which can be fixed by symmetry and Yang-Baxter -- see review \cite{chapSMat} and references therein -- and a scalar \textit{dressing factor} $\s^2$ which is the main focus of this review.

To avoid possible ambiguities let us mention that we use a definition of $\s$ such  that the Beisert-Staudacher Bethe equations \cite{Beisert:2005fw,Beisert:2005tm} in the $SU(2)$ sector read
$
    e^{ip_kL}=\prod_{j\neq k }\frac{u_k-u_j+i}{u_k-u_j-i}\,\s(u_k,u_j)^2.
$

In contradistinction with the $O(4)$ sigma model, the AdS/CFT integrable system \textit{does} have infinitely many bound states. The $n$-th bound state is a composite state of $n$ elementary magnons with rapidities \cite{Dorey:2006dq}
\beq
u+i j\,, \qquad \text{with} \qquad j=-\frac{n-1}{2} ,\dots, \frac{n-1}{2} \,,\la{rapstring}
\eeq
separated by $i$ and with real part $u$. The energy and momentum of the bound state are the sum of energies and momenta of the individual constituents with the rapidities (\ref{rapstring}). The result is again (\ref{dispersionAdS}) but with $\xpm$ replaced by $x\left(u\pm in/2\right)$. The existence of bound states is reflected in a simple pole of the scattering matrix at the point $u=v-i$. By convention, this pole is included in $\hat S_0$. Hence $\s^2(u,v)$ should be regular at this point.

As usual in integrable models, the most powerful tool to fix $\s^2(u,v)$ is by using crossing symmetry. The
AdS/CFT generalized crossing equation was first proposed by Janik \cite{Janik:2006dc} and is presented in (\ref{crossing}) below. As discussed in section \ref{O4sec}, one can motivate this relation by finding a singlet particle/antiparticle state $|1\rangle$ with zero charges and imposing trivial scattering of this object with physical particles \cite{Beisert:2005tm}. Yet another derivation of crossing is \cite{Arutyunov:2006yd}.

To be able to discuss the crossing equation we need to explain what the ${particle}\to{antiparticle}$ transformation is for the AdS/CFT system. This transformation should be understood as the following monodromy in $u$.
We start with some $x^{\pm}$ at real $u$. The energy and momentum following from (\ref{dispersionAdS}) are then real. Next, we take $u$ into the complex plane: first we cross the cut of $x^{+}(u)$ (in the lower half plane), then the cut of $x^-(u)$ (in the upper half plane) and finally we come back to the original value of $u$, see figure \ref{fig:gammacross}.\footnote{Another interesting transformation can be considered when we only cross the $x^{-}$ cut. If we now compute the energy and momenta from (\ref{dispersionAdS}) at real $u$ we see that they are purely imaginary. This is the so called mirror kinematics where we should identify $p_{mirror}=i E$ and $E_{mirror}=i p$, see \cite{Ambjorn:2005wa,Arutyunov:2007tc} and review \cite{chapTBA} for more details.} Of course, since we crossed cuts we are now at some other sheet; to stress the difference let us denote the point on the different sheet by ${\bf u}$.  Since we crossed the cuts of $x^{\pm}$ we have $x^{\pm}({\bf u})=1/x^{\pm}(u)$ and therefore $E({\bf u}),p({\bf u})=-E(u),-p(u)$ as expected for a particle to antiparticle transformation. In sum, we have,
\be\label{aparticleparticle}
    \xpm_{antiparticle}=\frac 1{\xpm_{particle}} \,.
\ee
Note however that it is important to specify the path $\gamma$ under which $x^{\pm}\to 1/x^{\pm}$ to properly define this transformation. The crossing relation then reads
\be\label{crossing}
  \sigma(u,v)\sigma^{\g}(u ,v)=\frac{1-\frac 1{\xp\yp}}{1-\frac \xm{\ym}}\frac{1-\frac \xm{\yp}}{1-\frac 1{\xp\ym}}\,,
\ee
where $\sigma^{\g}(u,v)$ means the analytical continuation of $\sigma(u,v)$ in the $u$ variable over the contour $\g$. We will use $x^{\pm}\equiv x^{\pm}(u)$ and $y^{\pm}\equiv x^{\pm}(v)$ in this review.
\subsubsection*{Assumptions on the analytical structure}
The crossing equation (\ref{crossing}) admits infinitely many solutions. To single out the correct one we need additional physically motivated constraints on the analytical structure of the dressing factor. This is the analogue of the minimality condition for the $O(4)$ case discussed in section \ref{O4sec}.
\bn
\item{\bf Bound states and simple poles/zeros.} The position of simple poles in the S-matrix should correctly reproduce the structure of bound states. The only bound states in the AdS/CFT system are the (\ref{rapstring}) described above. These states are already accounted for by the simple pole of $\hat S_0$. % at the position $u-v=i$.
We therefore require that $\s^2(u,v)$ does not have simple poles. Unitarity then excludes simple zeros as well.\footnote{In the $O(4)$ case we only required the absence of poles and zeros in the physical strip due to the periodicity properties of $p$ as a function of $\theta$. In the AdS/CFT case there is no such periodicity and we should require the absence of poles everywhere.}
\item{\bf Poles/zeros of higher degree.} An exceptional feature of $1+1d$ theories is that the S-matrix can contain poles of higher degree associated with multiparticle exchanges. In \cite{Dorey:2007xn} {Dorey, Hofman, and Maldacena showed} that the exchange of pairs of {composite} states should lead to double poles in the AdS/CFT scattering matrix at $u-v=im$ for integer $m$~\footnote{The possible values of $m$ are in general restricted depending on what Riemann sheets of $\s(u,v)$ we are located \cite{Dorey:2007xn}. We do not need this restriction; it will naturally come out.}. Since the matrix factor $\hat S_0$ does not contain such double poles we should allow for their presence in the dressing factor. These should be, however, the only poles of $\s^2(u,v)$.
Using unitarity, we see that the zeros of $\s^2(u,v)$ are double zeros located at $u-v=-im$. These double poles (and zeros) are called DHM poles (and zeros). Note that DHM poles will inevitably appear in the solution of crossing. Therefore we can reformulate our requirement as a demand to pick up solution with the simplest possible pole/zero structure.
     %The dressing factor does not have poles on the physical sheet where $|\xpm[u]|>1$. The dressing factor $\s(u,v)$ is of course just a small part of the AdS/CFT S-matrix. For example, the so called BDS factor \cite{Beisert:2004hm} -- which multiplies $\sigma^2$ in the $SU(2)$ graded Bethe equations -- already contains poles in the physical sheet associated to the BPS bound-states of the theory \cite{Dorey:2006dq}, see also review \cite{chapTBA}.
%At weak and strong coupling -- when the dressing factor can be explicitly computed -- no other poles arise and hence this is a reasonable property to assume at finite coupling, it is the analogue of the minimal choice condition we saw in the $O(4)$ sigma model example.
%
% It is usually argued \cite{}(Arutyunov) that physically meaningful solutions of the Beisert-Staudacher Bethe Ansatz equations satisfy the condition $|\xpm[u]|>1$. The main reason for this (*are there any other reasons??*) is that the energy of only these solutions has a nonsingular $g\to 0$ limit. Due to this the Riemann sheet $|\xpm[u]|>1$ of the $u$ plane is identified with the physical sheet. The small coupling limit of the dressing factor $\s[u,v]$ is equal to $1$ and therefore does not contain poles. Based on this we assume that $\s[u,v]$ does not have poles on the physical sheet for finite $g$ as well.
%
%    The absence of poles For ${\rm Im}[u]<0$ is an analog of the minimality condition in the relativistic theories. (And what problems with poles for ${\rm Im}[u]>0$??)
\item{\bf Branch points.} In the $O(4)$ sigma model, the S-matrix, when thought of as a function of the kinematic invariants, has branch cut singularities at particle creation thresholds. The rapidity $\theta$ uniformizes the $S$-matrix rendering it meromorphic.
%as a function of the Mandelstam variable $s$ has branch points at $s=0$ and $s=4m^2$. The rapidity $\theta$ uniformizes
%
%An important step in the solution of the crossing equation is the requirement that in terms of the rapidity variable $\theta$ the S-matrix becomes a meromorphic function.
%
In the AdS/CFT system we are not able to introduce such uniformizing variable. The best we can do is to require that the structure of branch points is as simple as possible: only the branch points which are explicitly required by crossing equations are allowed in $\sigma(u,v)$. We will see that this criterium leads to infinitely many square-root branch points at the points $u=\pm 2g \pm in$ for half-integer $n$, i.e. when $x(u\mp in)=\pm 1$.
% Recall that the Zhukovsky variables $x(u\pm in/2)$ naturally parametrize the energy and momenta of bound-states made out of $n$ fundamental particles. The two-body S-matrix should contain information about all these bound states so the structure of branch points of $\sigma^2(u,v)$ is not that surprising.

\item{\bf $\chi$-decomposition} The dressing factor  $\s(u,v)=e^{i\theta(u,v)}$ can be decomposed as 
\be\label{parame}
\theta(u,v)=\chi(\xp,\ym)-\chi(\xm,\ym)-\chi(\xp,\yp)+\chi(\xm,\yp),
\ee
where $\chi(x,y)$ is antisymmetric, $\chi(x,y)=-\chi(y,x)$, and $\theta(u,v)$ is the \textit{phase shift}. This form
is the most general expression we should expect for long-range integrable spin chains \cite{Beisert:2005wv}. Hence, from the $\mathcal{N}=4$ point of view it is perfectly justified.  More precisely this form is a direct consequence of the decomposition of $\theta(u,v)$ in terms of higher conserved charges (\ref{expansionsigma}) which will be reviewed in more detail in section \ref{sec:urae}. The higher charge decomposition property was realized in \cite{Arutyunov:2004vx,Beisert:2005wv}, while the representation (\ref{parame}) appeared in \cite{Arutyunov:2006iu}. Further evidence for this decomposition comes from considering the scattering of bound-states, see point $6$ below.

%  \item{\bf $\chi$-decomposition} The dressing factor can be  represented in the form
%  \cite{Arutyunov:2004vx,Beisert:2005wv}:
%  \be\label{parame}
%  \s(u,v)=e^{i\theta(u,v)},\ \
%  \theta(u,v)=\chi(\xp,\ym)-\chi(\xm,\ym)-\chi(\xp,\yp)+\chi(\xm,\yp),
%  \ee
%  where $\chi(x,y)$ is antisymmetric, $\chi(x,y)=-\chi(y,x)$, and $\theta(u,v)$ is the \textit{phase shift}.
%  This form is the most general expression we should expect for long-range integrable spin chains
%  \cite{Beisert:2005wv}. Hence, from the $\mathcal{N}=4$ SYM point of view, this ansatz is perfectly justified.

\item{\bf Asymptotics at infinity.} As $x\to \infty$ we expect $\chi(x,y)$ to approach some constant value. Since infinite $x$ corresponds to zero momenta this will ensure that $\sigma^2(p=0,p') \to 1$, i.e. particles scatter trivially with zero momentum particles. This should be imposed since excitations with zero momenta correspond to global symmetry transformations of the state and should therefore have an irrelevant effect.

\item{\bf Analyticity of  $\chi(x,y)$ in the physical domain $|x|>1$.}
In points 2 and 3 above we anticipated the existence of DHM poles and of square root singularities at $u=v+ in$ and $u=\pm 2g \pm in$ respectively. So, what we want to argue in this point is that these singularities should not be present in the physical sheet $|x(u)|>1$. This is the trickiest point and requires a somehow more involved argument. The basic idea is that if these singularities would be present for $|x|>1$ then they would lead to unphysical singularities in the description of the scattering of bound states. To show this we have to consider the phase shift when scattering a $n$-th with a $m$-th bound state with real rapidities $u$ and $v$. The total phase shift is the sum of the phase shifts acquired by each of the $n$ constituents of the first bound state when scattered through each of the $m$ constituents of the second bound state. I.e. $    \theta_{n,m}(u,v)=\sum_{j=-\frac {n-1}2}^{\frac{n-1}2}\sum_{k=-\frac{m-1}2}^{\frac{m-1}2}\theta(u+i\,j,u+i\,k)
$. Using (\ref{parame}) we simplify this sum to \cite{Chen:2006gq}\footnote{The nice fusion properties of the dressing factor can be seen as further evidence in favor of the composition (\ref{parame})}
\be
    \theta_{n,m}(u,v)&=&\chi(x(u+in/2),x(v+im/2))-\chi(x(u-in/2),x(v+im/2))\no\\
    &&-\chi(x(u+in/2),x(v-im/2))+\chi(x(u-in/2),x(v-im/2)). \no\\%\no
\ee
The scattering matrix $\hat S(u,v)$ should be analytic for real $u$ and $v$ provided we are in the physical domain $|x(u\pm in/2)|>1$. If the square root cuts or the DHM poles were present in the physical region of $\chi(x,y)$ they would clearly lead to singularities in the real axis for some $n$-th bound states  (of even n). Hence they must be absent.

%
%This is a natural physical condition since we expect scattering to be regular process.
%
%The consequence of this requirement is that $\chi(x,y)$ should be analytic in the region $|x|>1$. Indeed, $\chi(x,y)$ may have two kind of singularities: branch points at $x=x(\pm 2g \pm in)$ and DHM poles at $x=x(v+ in)$ for $n$ being positive integer. But if any of these singularities is present in the physical domain, the scattering matrix of the $2n$-th bound state would contain this singularity shifted to the real axis, which is forbidden.

%The reason that the decomposition (\ref{parame}) is possible is that for at least sufficiently large $u$ and $v$ $\theta[u,v]$ is decomposable in terms of the conserved charges:
%\be\label{expansioncharges}
%    \theta[u,v]&=&\sum_{r,s}c_{r,s}Q_{r}[x]Q_s[y],\no\\
%    Q_r[x]&=&\frac 1{\left( \xp\right)^{r-1}}-\frac 1{\left( \xm\right)^{r-1}}
%\ee
%of the previous point  Assumption on the analyticity of the dressing factor makes (\ref{expansioncharges}) convergent and leads to (\ref{param}). The minimality condition leads to the requirement

%This asymptotics is compatible with the asymptotics of the strong coupling expansion of the dressing phase which is known from string theory calculations.
\en
\subsubsection*{Solution}
In this derivation we follow \cite{Volin:2009uv} closely. The crossing relation (\ref{crossing}) is valid on the infinite genus Riemman surface {\cite{Beisert:2006ib}} where $u$ lives. Instead of evaluating it at a real $u$ in the physical sheet where $|x^{\pm}(u)|>1$ let us cross the cut of $x^{-}(u)$ and return back to the real $u$ axis. This means that we should flip $x^-\to 1/x^-$ in the right hand side of (\ref{crossing}) which becomes\footnote{This more symmetric form of crossing is the analogue of (\ref{crossingO4}) where we evaluate the dressing factors at $\theta\pm i/2$, see comment at the end of section \ref{O4sec}. Similarly to what happens in the $O(4)$ model, the derivation also simplifies when we consider this more symmetric form.}
\be\label{mirrorcros}
  \s^{\g_-}(u,v)\s^{\g_+}(u,v)=\frac{1-\frac 1{\xp\yp}}{1-\frac 1{\xm\ym}}\frac{1-\frac 1{\xm\yp}}{1-\frac 1{\xp\ym}},
\ee
where $\s^{\gamma_{\pm }}(u,v)$ is the analytical continuation of $\s(u,v)$  through the $u$ contour which crosses the cut of $x^{\pm}$ so that $x^{\pm} \to 1/x^{\pm}$ respectively, see figure \ref{fig:gammacross}. Of course, the contour $\gamma$ described above (\ref{crossing}) is nothing but $\gamma=\gamma^+ + \(\gamma^-\)^{-1}$.
Next we notice that $\sigma^{\g_-}=\frac {\s_1(\xp,v)}{\s_1(1/\xm,v)}$ and $\sigma^{\g_+}=\frac {\s_1(1/\xp,v)}{\s_1(\xm,v)}$
%\be\nn
%  \sigma^{\g_-}=\frac {\s_1(\xp,v)}{\s_1(1/\xm,v)},\ \
%  \sigma^{\g_+}=\frac {\s_1(1/\xp,v)}{\s_1(\xm,v)}.
%\ee
where $\s_1(x,v) \equiv e^{i\chi(x,\ym)-i\chi(x,\yp)}$, see decomoposition (\ref{parame}). By plugging these expressions into the crossing relations we get
\beq
\frac{\sigma_1(x^+,v) \sigma_1(1/x^+,v)}{\sigma_1(x^-,v) \sigma_1(1/x^-,v)} = \frac{1-\frac 1{\xp\yp}}{1-\frac 1{\xm\ym}}\frac{1-\frac 1{\xm\yp}}{1-\frac 1{\xp\ym}} \la{crossing2} \,.
\eeq
As in the $O(4)$ sigma model, it is useful to manipulate expressions by using the shift operator $D=e^{\frac{i}{2}\partial_u}$. The shorthand notation for (\ref{crossing2}) reads
\be\label{mirror}
  &&\( \s_1(x,v)\s_1(1/x,v) \)^{D-D^{-1}}=\(\frac{x-\frac 1{\yp}}{x-\frac 1{\ym}}\)^{\!\!D+D^{-1}}.
\ee
The use of shift operators in the AdS/CFT system is potentially dangerous due to the presence of the branch points so let us analyse expression (\ref{mirror}) with care.
The right hand side of (\ref{mirror}) is uniquely defined in the region $|{\rm Re}(u)|>2g$ if we use the physical choice of cuts for $x(u)$, see figure \ref{fig:gammacross}. The reason is that if $|{\rm Re}(u)|>2g$ we never cross the cuts of $x^{+}(u)$ or $x^{-}(u)$.
Interestingly, the left hand side of (\ref{mirror}) is not ambiguous at all in the whole strip $|{\rm Im}(u)|<1/2$, and in particular on the real axis. Indeed, when we cross the cut of $x(u)$, the two terms in the product $\s_1(x,v)\s_1(1/x,v)$ become just exchanged, so this product does not have the cut of $x(u)$.\footnote{Of course, this product might still have cuts of $x(u+in)$ with nonzero integer $n$ and this is why we restrict ourselves to the strip $|{\rm Im}(u)|<1/2$ to be on the safe side.}
We conclude that it is safe to use the shift operator for the expression (\ref{mirror}) at least in the intersection of two domains $|{\rm Re}(u)|>2g$ and $|{\rm Im}(u)|<1/2$. We therefore consider (\ref{mirror}) in this intersection, solve it,  and then analytically continue the solution everywhere.

Formally we can solve (\ref{mirror}) by\footnote{A priori we could imagine multiplying the right hand side of (\ref{good}) by a zero mode of $D-D^{-1}$, i.e. a function $g(u)$ periodic in $u$ with period $i$. Such functions must however always have poles or zeros.  As explained in the previous section, the only allowed poles (zeros) are the DHM poles (zeros). Suppose that $g(u)$ contains any of DHM poles. Then by periodicity it contains poles at $u=i\MZ$. But due to unitarity it contains also zeroes in the same positions. Hence $g(u)$ is a constant which can be set to $1$ due to (\ref{parame}). Hence  (\ref{good}).}
\be\label{s1}
  \s_1(x,v)\s_1(1/x,v)=\(\frac{x-\frac 1{\yp}}{x-\frac 1\ym}\)^{f[D]} \,,\la{good}
\ee
where \beq f[D]=\frac{D+D^{-1}}{D-D^{-1}}\,.\eeq However, to interpret this expression we must give a meaning to this operator.\footnote{This can be almost xeroxed from the discussion of the dressing factor in the $O(4)$ sigma model in section \ref{O4sec} where we needed to regularize a very similar expression, see (\ref{sigma1}).}
For example, a naive expansion in powers of $D$ or $1/D$ leads to $f[D]=-1 + \mathcal{O}(D)$ or $f[D]=+1 + \mathcal{O}(1/D)$ respectively. Both expansions must be discarded, because of the $\mp 1$ terms: the presence of such terms would mean that $\s_1(x,v)\s_1(1/x,v)$ has the branch-cuts of $x(u)$, however this should not be the case as explained above.  The proper interpretation of $f[D]$ which leads to a function $\s_1(x,v)\s_1(1/x,v)$ without a branch cut for $u\in [-2g,2g]$ is given by\footnote{Strictly speaking, the expression  (\ref{s1}) with $f[D]$ given by (\ref{goodf}) still needs to be regularized to have a precise meaning. For instance, instead of $f[D]$ we might consider $\pd_u^2 f[D]$ and then integrate back. However, the terms that depend on the regularization are canceled out when computing the dressing factor $\s(u,v)$ because of the anti-symmetrization (\ref{parame}) over the several $\chi$ factors.
}
\beq
f[D]=\frac {D^{-2}}{1-D^{-2}}-\frac{D^2}{1-D^2}\equiv \sum_{n=1}^{\infty} D^{-2n}- \sum_{n=1}^{\infty}D^{2n} \, .\la{goodf}
\eeq
The reason is of course the absence of $D^0$ terms in this expansion. Plugging the definition of  $\s_1(x,v)$ in terms of $\chi(x,y)$ in (\ref{good}) we see that it can be further factorized into
\be
  e^{i\chi(x,y)+i\chi(1/x,y)}=\(\frac{x-\frac 1y}{\sqrt{x}}\)^{-f[D]}\,. \la{goodchi}
\ee
The factor $1/\sqrt{x}$ is irrelevant for $\s_1$ but we insert it to ensure the antisymmetry of $\chi(x,y)$ with respect to the interchange $x\leftrightarrow y$. A direct calculation yields\footnote{the logarithm of the right hand side is antisymmetric with respect to $u\leftrightarrow v$ as it should.}
\be\label{s2}
  e^{i\chi(x,y)+i\chi(1/x,y)+i\chi(x,1/y)+i\chi(1/x,1/y)}=(u-v)^{-f[D]}=\frac{\Gamma(1+iu-iv)}{\Gamma(1-i u+i v)} \,.
\ee
Now, if we take $u$ on top of the cut of $x(u)$ we have $x(u-i0)=1/x(u+i0)$. Similarly for $v$. Hence, with an harmless abuse of notation, we can think of (\ref{s2}) as a Riemann-Hilbert problem:
\beq%\nn
\chi(u\!+\!i0,v\!+\!i0)\!+\!\chi(u\!-\!i0,v\!+\!i0)\!+\!\chi(u\!+\!i0,v\!-\!i0)\!+\!\chi(u\!-\!i0,v\!-\!i0)=\frac 1i\log \frac{\Gamma(1+iu-iv)}{\Gamma(1-iu+iv)}.
\eeq
Needless to say, Riemann-Hilbert problems are much simpler than generic functional equations -- such as the original crossing equation --  and can be solved by standard methods. In our case the solution is given by
\be\label{s2solution}
  \chi(x,y)=\frac{1}{i} {K}_u \star {K}_v \star \log \frac{\Gamma(1+iu-iv)}{\Gamma(1-iu+iv)},
  \ee
with the kernel $K$ defined as\footnote{Note that $x(u)-1/x(u)=g^{-1}\sqrt{u^2-4g^2}$, so $K_u$ is the typical kernel used to solve Riemann-Hilbert problems of the form (\ref{Cauchy}).}
\be\label{Ktildedef}
  K_u\star F \equiv\int\limits_{-2g+i0}^{2g+i0}\frac{dw}{2\pi i}\,\,\frac {x(u)-\frac 1{x(u)}}{x(w)-\frac 1{x(w)}}\,\frac 1{w-u}\,F(w)\,.
\ee
The kernel  is engineered to satisfy the following equation\footnote{This kernel coincides, after an analytical transformation \cite{Kostov:2008ax}, with the inverse Fourier transform of the sum $K_0+K_1$ of the magic kernels in \cite{Beisert:2006ez}. The relevance of Riemann-Hilbert problem for $K_u$ was recognized in \cite{Kostov:2008ax}.}:
\be\label{Cauchy}
  (K_u\star F)(u+i0)+(K_u \star F)(u-i0)=F(u) \,\, ,\qquad  |u|<2g.
\ee
The solution (\ref{s2solution}) was chosen among the other possible solutions by the requirement that $\chi(x,y)$ should be analytic for $|x|>1$ and $\chi(x,y)\to{\rm const}$ as $x\to\infty$. The expression (\ref{s2solution}) can be rewritten in the form proposed in \cite{Dorey:2007xn} if we rewrite the action of the kernels as an integral in the Zhukovsky plane,
\be\label{toDHM}
  (K_u \star F)(u)=\int\limits_{|z|=1}\hspace{-1.30EM}\circlearrowleft\hspace{1EM}\frac {dz}{2\pi i}\frac 1{x-z}F\(g\(z+{1}/{z}\)\)-\frac{1}{g}\int\limits_{-2g+i0}^{2g+i0}\frac{dv}{2\pi i}\frac{  1}{x(v)-\frac 1{x(v)}} F(v)\,,
\ee
and notice that the second term does not contribute to the dressing phase once we anti-symmetrize in (\ref{parame}). Dropping it, we obtain the DHM representation
\be\label{DHM}
%&&\s[u,v]=e^{i\theta[u,v]},\ \ \theta[u,v]=(D_u-D^{-1}_u)(D_v-D^{-1}_v)\chi[x(u),x(v)],\no\\
\chi(x,y)=-i\int\limits_{|z|=1}\hspace{-1.33EM}\circlearrowleft\hspace{1EM}\frac {dz}{2\pi i}\int\limits_{|z'|=1}\hspace{-1.43EM}\circlearrowleft\hspace{1EM}\frac {dz'}{2\pi i}\frac 1{x-z}\frac 1{y-z'}\log\frac{\Gamma[1+ig(z+\frac 1z-z'-\frac 1{z'})]}{\Gamma[1-ig(z+\frac 1z-z'-\frac 1{z'})]} \,.
\ee

\subsubsection*{Analytical structure at finite coupling}
%The investigation of the analytical structure of the solution (\ref{s2solution}) is based on the property (\ref{Cauchy}). It is instructive to write (\ref{s2solution}) as
\begin{figure}[t]
\centering
\includegraphics[width=13.00cm]{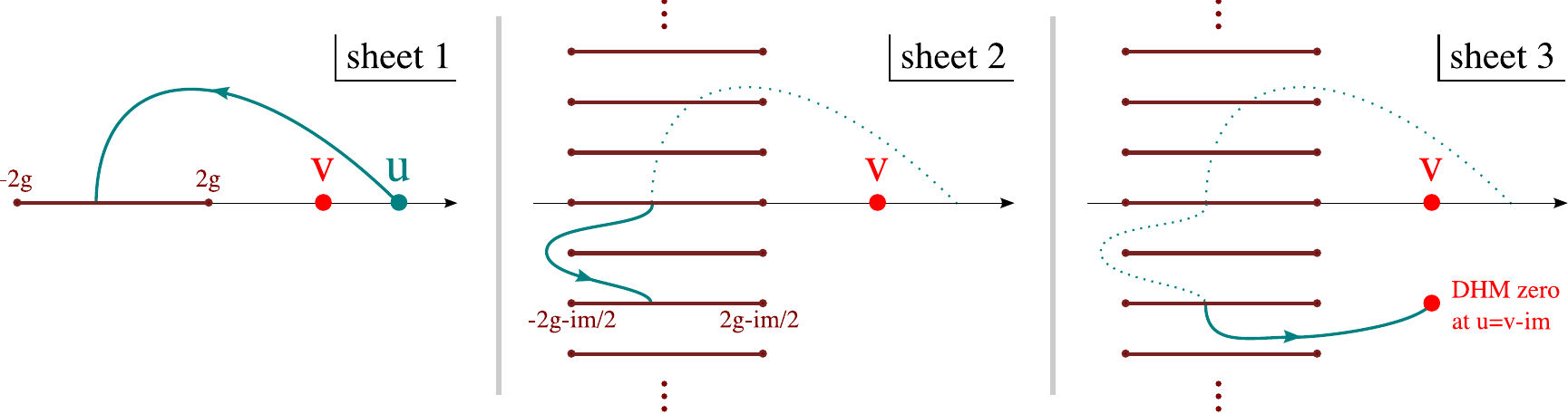}
\caption{Analytical structure of $e^{i\chi(x,y)}$ as a function of $u$.} \label{fig:dressinganalytics}
\end{figure}

To investigate the analytical structure of $\chi(x(u),x(v))$ as a function of $u$ we will use (\ref{goodf}) to write\footnote{The sharp reader will have noticed that $K_v$ was replaced by $K_u$. Such replacements are allowed since the difference between both expressions cancels out in the anti-symmetrization (\ref{parame}). The (derivative of the logarithm of the) representation (\ref{repre}) for the dressing phase was proposed in \cite{Kostov:2007kx}.}:
\be\label{repre}
i\chi(x(u),x(v))=\sum_{n\neq 0}{\sign\!(n)}  K_u \star  D^{2n} K_u \star  \log(u-v)
\ee
which is valid in the physical domain $|x(u)|>1$. To go to $|x(u)|<1$ we must cross the cut $u \in [-2g,2g]$. Using (\ref{Cauchy}) to go through this cut, we obtain
\beq
\chi(x(u),x(v))\!=\!i\!\sum_{n\neq 0}{\sign\!(n)}  K_u \star  D^{2n} K_u \star  \log(u-v) -i\!\sum_{n\neq 0}\sign\!(n) D^{2n} K_u \star \log(u-v). \la{second}
\eeq
%\beqa
%i\chi(x(u),x(v))&=&-\sum_{n\neq 0}{\sign\!(n)}\  K_u \star  D^{2n} K_u \star  \log(u-v) \nn\\
%&&+\,\,\sum_{n\neq 0}\sign\!(n)\, D^{2n} K_u \star \log(u-v). \la{second}
%\eeqa
Since the only cut we crossed was the cut of $x(u)$, this expression is valid in the domain where $|x(u)|<1$ while $|x[u+in]|>1$ for $n\neq 0$. By crossing the cut of $x(u)$ we moved into a different Riemann sheet. On this new sheet $\chi(x(u),x(v))$ has all the cuts of $x(u+in)$ for $n \in \mathbb{Z}$ as depicted in figure \ref{fig:dressinganalytics}. We could now decide to go through one of the new cuts, e.g. $|x(u+im)|=1$ with $m\neq 0$. This will bring us to a third Riemann sheet which is now defined by $|x(u)|<1$ and $|x(u+im)|<1$ with all other $|x(u+in)|>1$. When going through the cut of $|x[u+im]$ we pick an extra contribution from the second term in (\ref{second}) so that
\beqa
i\chi(x(u),x(v))\!\!\!\!\!&=&\!\!\!\!\!-\sum_{n\neq 0}{\sign\!(n)}\  K_u \star  D^{2n} K_u \star  \log(u-v) \nn+\sum_{n\neq 0\!,\,m}\sign\!(n)\, D^{2n} K_u \star \log(u-v)\nn\\
&&\!\!\!\!\!\!\!\!\!\!-\ \sign\!(m)\ D^{2m} K_u \star \log(u-v)+\,\sign\!(m)\log(u-v+i m).\la{third}
\eeqa
%\beqa
%i\chi(x(u),x(v))&=&-\sum_{n\neq 0}{\sign\!(n)}\  K_u \star  D^{2n} K_u \star  \log(u-v) \nn\\
%&&+\sum_{n\neq 0\!,\,m}\sign\!(n)\, D^{2n} K_u \star \log(u-v)\nn\\
%&&-\ \sign\!(m)\ D^{2m} K_u \star \log(u-v)+\,\sign\!(m)\log(u-v+i m).\la{third}
%\eeqa
When constructing $\s^2$ the last term leads to double poles or double zeros (depending on the sign of $m$). These are precisely the DHM poles/zeros mentioned above.
This is a very nontrivial \textit{finite coupling} check of the dressing factor.

\section{Useful representations and expansions\label{sec:urae}} \la{expsec}
A particularly useful alternative way of expressing $\chi(x,y)$ is through its large $x$ and $y$ expansion,
\beq
\chi(x,y)=-\sum_{r,s=1}^{\infty} \frac{c_{r,s}(g)}{x^{r}y^s}\,, \la{sumform}
\eeq
where $c_{r,s}=-c_{s,r}$ are a set of functions of the 't Hooft coupling only. For the dressing factor this translates into an expansion in terms of magnon higher conserved charges:
\beq
\frac{1}{i}\log \sigma(u,v)= \sum_{r,s=1}^{\infty} c_{r,s}(g) \,q_{r+1}(u)\,q_{s+1}(v) \,,\,\,\,\, \qquad \,\,\,\,\, q_{r}(u)\equiv \frac{1}{\(x^+\)^{r-1}}-\frac{1}{\(x^-\)^{r-1}}\,. \la{expansionsigma}
\eeq
The higher conserved charges $q_r(u)$ were initially written in \cite{Beisert:2004hm} and the expansion (\ref{expansionsigma}) was proposed in \cite{Arutyunov:2004vx} (up to minor modifications). This proposal found further support in \cite{Beisert:2005wv} where Beisert and Klose argued that generic $gl(r)$ symmetric long-ranged integrable spin chain models give rise to (\ref{expansionsigma}).\footnote{For more recent discussions of general integrable long range Hamiltonians see \cite{chapLR}.} Expanding (\ref{DHM}) at large $x$ and $y$ and parameterizing $z,z'=e^{\phi},e^{\phi'}$, we find
\beq
c_{r,s}(g)=i\int\limits_{0}^{2\pi} \frac{d\phi}{2\pi} \int\limits_{0}^{2\pi} \frac{d\phi'}{2\pi} \exp\(i r \phi + i s \phi'\)  \log\frac{\Gamma[1+2ig( \cos \phi - \cos \phi')]}{\Gamma[1-2ig( \cos \phi - \cos \phi')]}. \la{DHMcs}
\eeq
The logarithm of gamma functions in this expression has an integral representation as $ -2i\, {\rm Im} \int_{0}^{\infty} \frac{dt}{t}\[ \frac{e^{2igt(\cos \phi- \cos \phi')}-1}{e^{t}-1}-2ig\(\cos\phi-\cos \phi'\) e^{-t} \]$.
Only the exponential term survives after the $\phi$ and $\phi'$ integration. Furthermore, these angular integrals yield integral representations of Bessel functions. Hence,
\beq
c_{r,s}(g)= 2 \sin\( \frac{\pi}{2}(r-s)\)  \int\limits_{0}^{\infty} dt \, \frac{J_{r}(2gt)J_{s}(2gt)}{t(e^t-1)} \, . \la{BEScs}
\eeq
Our logical flow was pretty much the exact opposite of the chronological one. The dressing factor was first guessed by Beisert-Eden-Staudacher \cite{Beisert:2006ez} in the form (\ref{sumform}),(\ref{BEScs}) based on the string analysis of Beisert-Hernandez-Lopez \cite{Beisert:2006ib} and on trancendentality considerations \cite{Kotikov:2002ab}. Dorey-Hofman-Maldacena \cite{Dorey:2007xn} derived (\ref{DHMcs}) from (\ref{BEScs}) and resumed (\ref{sumform}) to derive the integral representation (\ref{DHM}). Only later the dressing factor was shown to {satisfy the crossing equation in \cite{Arutyunov:2009kf} and explicitly derived in \cite{Volin:2009uv}.}

\subsubsection*{Weak coupling expansion}
The constants $c_{r,s}(g)$ admit a convergent weak coupling expansion as
\beq
c_{r,s}(g)=\sum_{n=0}^{\infty} c^{(n)}_{r,s} \, g^{r+s+2n}  \la{weak}
\eeq
where \cite{Beisert:2006ez}
\beq
c_{r,s}^{(n)}=2 \,(-1)^{n}   \sin\( \frac{\pi}{2}(r-s)\)  \frac{(2 n+r+s-1)! (2 n+r+s)! }{n! (n+r)! (n+s)! (n+r+s)!} \zeta (2 n+r+s) \la{BEScsg}\,.
\eeq
The convergence radius of this expansion is $g_c=1/4$, see figure \ref{convergence}. A simple explanation for this radius of convergence is that at $g=i/4$ the branch points of the dressing factor collide in pairs. As one can see from (\ref{BEScsg}), the constants $c_{r,s}(g)$ behave at weak coupling as
\be\label{order}
   c_{r,s}(g)=\CO(g^{r+s}).
\ee
This was predicted in \cite{Beisert:2005wv} as the generic behavior of the constants $c_{r,s}(g)$ for spin chain arising from perturbative computations in gauge theories in the planar limit. It is therefore a very important check of the solution (\ref{DHM}).

The leading weak coupling term is $c_{1,2}=-c_{2,1}=-2g^{3} \zeta(3)$. For rapidities  $u,v=\mathcal{O}(1)$ it leads to
\beq
\sigma^2(u,v)=1+256\, \zeta(3) g^6\,\frac{(u-v)(4uv-1) }{(1+4u^2)^2(1+4v^2)^2} +\mathcal{O}(g^8) \,. \la{expansion}
\eeq
This means that the effect of the dressing factor only affects the rapidities of the particles at order $g^6$. The energy of a multi-particle state is given by a sum of dispersion relations $\sum_{j} \epsilon(u_j)$ where $\epsilon(u_j)=\mathcal{O}(g^2)$ which means that the dressing factor only affect the anomalous dimensions of single trace operators at order $g^{8}$, i.e. at $4$ loops!  It is therefore no surprise that it was originally thought that such factor was absent all together \cite{Beisert:2004hm}.

The cusp anomalous dimension\footnote{Operators of the form $\Tr \(Z D^S Z\)+\text{permutations}$ have an anomalous dimension $\Delta(S,g)\simeq f(g) \log S $ at large $S$. The cusp anomalous dimension is $f(g)/2$.}, for example, can be computed using the dressing factor derived above \cite{Eden:2006rx,Beisert:2006ez}. One finds

\begin{figure}[t]
\centering
\includegraphics[width=6.00cm]{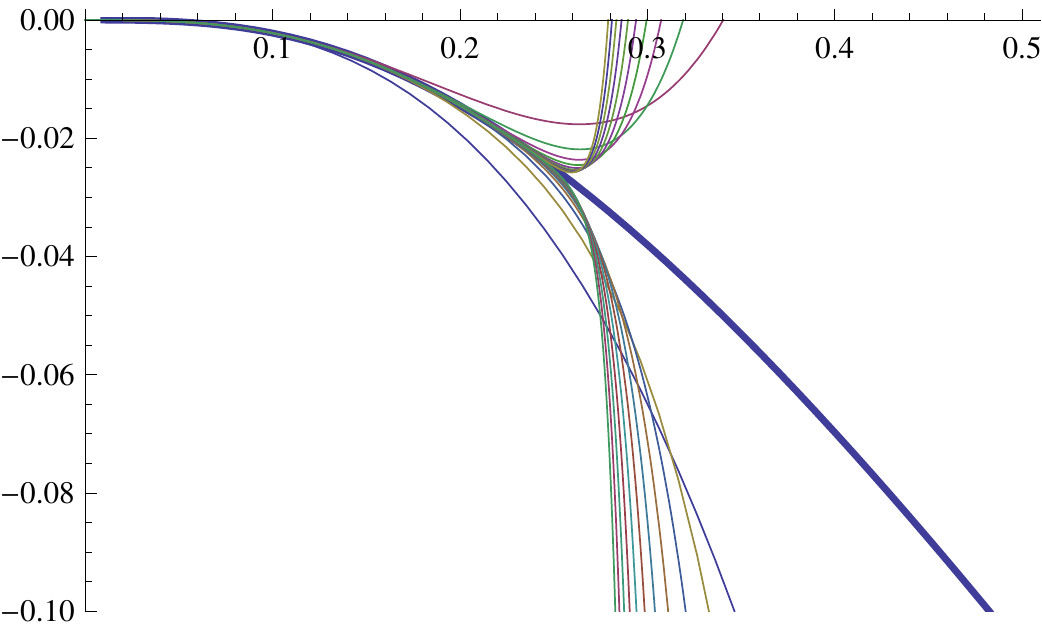}\ \ \
\includegraphics[width=6.00cm]{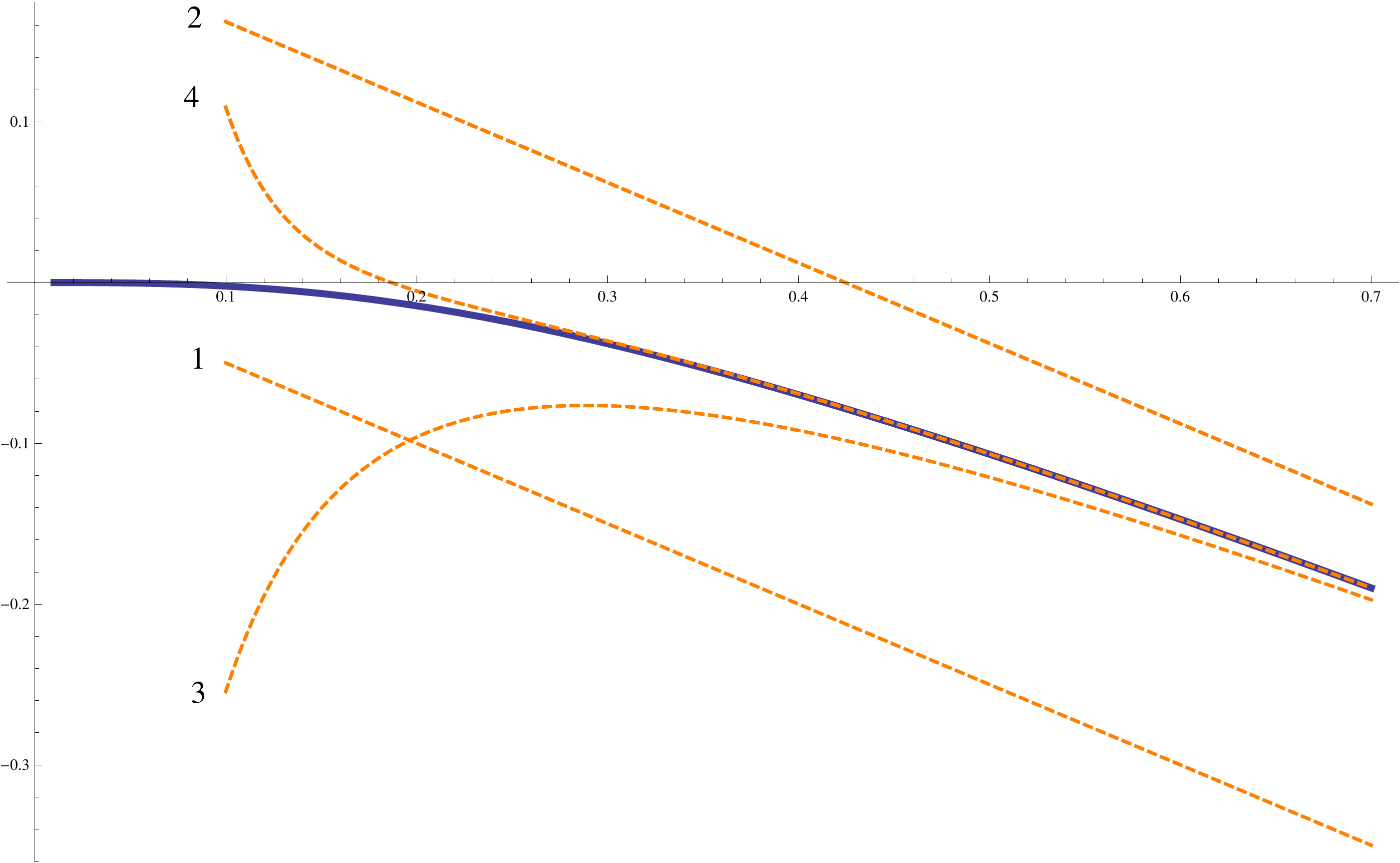}
\caption{ {\bf Left:}
$c_{1,2}(g)$ as given by (\ref{DHMcs}) or (\ref{BEScs}) can be evaluated for any $g$: thick line. The weak coupling Taylor expansion (\ref{weak}) has a finite radius of convergence. The several thin lines plotted in this figure are several truncations of this expansion (from $1$ to $20$ terms); the more terms are included the most they approach the real curve for $g<1/4$, the radius of convergence. {\bf Right:}
$c_{1,2}(g)$ (blue line) compared with its strong coupling expansion (dotted orange lines). $1$ is the AFS result, $2$ is AFS plus HL, $3$ already includes the two loop correction and $4$ contains the first four non-trivial terms. We see that the fit with four terms is perfect even for relatively small $g$! This fast convergency was also observed in numerical computations of dimensions, %Indeed, it has been observed in numerical computations that the strong coupling limit is reached for relatively small couplings,
see e.g. \cite{Benna:2006nd}.} \la{convergence}
%\parbox[t]{0.48\textwidth}{\caption{\footnotesize{\label{strong}%\textbf{Right:}
%Coefficient $c_{1,2}(g)$ (solid thick blue line) compared with its strong coupling expansion (dotted orange lines). The number close to each dotted line indicates how many terms were used to plot the function: $1$ corresponds to the AFS result, $2$ to AFS plus HL, $3$ already includes the two loop correction and $4$ contains the first four non-trivial terms of the strong coupling expansion. With only four terms of the strong coupling expansion we see that the fit to the exact coefficient is perfect even for relatively small $g$! Indeed, it has been observed in numerical computations that the strong coupling limit is reached for relatively small couplings, see e.g. \cite{Benna:2006nd,Gromov:2009zb}.}}}
\end{figure}
%\begin{figure}[t]
%\centering
%\includegraphics[width=8.00cm]{convergent}
%\caption{\footnotesize{\label{convergence} $c_{1,2}(g)$ as given by (\ref{DHMcs}) or (\ref{BEScs}) can be evaluated for any $g$: thick line. The weak coupling Taylor expansion (\ref{weak}) has a finite radius of convergence. The several thin lines plotted in this figure are several truncations of this expansion (from $1$ to $20$ terms); the more terms are included the most they approach the real curve for $g<1/4$, the radius of convergence.} }
%\end{figure}
\beqa
f(g)=8g^2-\frac{8\pi^2}{3} g^4+\frac{88 \pi^4}{45} g^2 - \(\frac{315\pi^6}{584}+64\zeta(3)^2\)g^8\nn \\
+\(\frac{28384\pi^8}{14175} +\frac{128\pi^2 \zeta(3)^2}{3} +1280 \zeta(3)\zeta(5)\)g^{10}-\dots \,. \la{cusp}
\eeqa
This expansion deserves a couple of comments. First notice that the degree of transcendentally of each term is correlated to the corresponding order of perturbation theory. This is in perfect agreement with the Kotikov-Lipatov transcendentality conjecture \cite{Kotikov:2002ab}. The zeta functions in (\ref{BEScsg}) are precisely of the required degree not to spoil this nice property! Second, if we were to compute the cusp anomalous dimension using $\sigma^2(u,v)=1$, we would find \textit{exactly the same result (\ref{cusp}) with the replacement $\zeta(2n+1)\to i \zeta(2n+1)$}! Quite mysteriously, the constants $c_{r,s}^{(n)}$ are \textit{uniquely} fixed to (\ref{BEScsg}) provided (a) we assume (\ref{order}) and (b) require that the presence of the dressing factor simply amounts to $\zeta(2n+1)\to i \zeta(2n+1)$ in the cusp anomalous dimension computed without any dressing factor.

The cusp anomalous dimension turns out to be a very important quantity in the study of gluon scattering amplitudes in $\mathcal{N}=4$ SYM \cite{Korchemsky:1991zp}. It governs the IR divergent part of these amplitudes. The cusp anomalous dimension has been computed analytically up to three loops \cite{Vogt:2004gi} and numerically at four loops \cite{Bern:2006ew}. Thus the first line in (\ref{cusp}) can be checked: it matches precisely the perturbative calculations!

Let us mention two more remarkable weak coupling checks of the dressing factor. At weak coupling it is convenient to think of operators
\beq
O(x)=\Tr\(Z\dots Z X Z \dots Z \Psi Z \dots Z DZ\dots Z \)(x)  + \text{permutations} \la{operator}
\eeq
as spin excitations $X,\Psi,D,\dots$ around a ferromagnetic spin chain vacuum $\Tr \,Z^L$. In this language the anomalous dimension matrix can be thought of as a spin chain Hamiltonian \cite{chapChain}. At four loops we have a Hamiltonian of range four explicitly computed in \cite{Beisert:2007hz}. At this loop order some coefficients of this Hamiltonian get $\zeta(3)$ factors. These lead precisely to the dressing factor (\ref{expansion})!

Other impressive weak coupling checks concerns the computation of the anomalous dimension of short operators such as the Konishi operator. At four loops the range of interaction of the Hamiltonian is as large as the operator itself and the scattering picture breaks down \cite{Staudacher:2004tk}. Still, using the Luscher formalism, this correction can be computed \cite{chapLuescher}. This prediction was checked against a tour de force computation \cite{Fiamberti:2008sh} and agreement was found. Other remarkable four and five loop checks concern the behavior of general twist two operators as predicted from the AdS/CFT system with the predictions of BFKL, see reviews \cite{chapTwist} and \cite{chapLuescher} for details.

\subsubsection*{Strong coupling}
The dressing factor can also be expanded at strong coupling. However, contrary to the expansion at weak coupling which was convergent, at strong coupling the expansion is merely asymptotic, albeit Borel summable. We have 
$c_{r,s}(g)= \sum_{n=0}^{\infty} \,d_{r,s}^{\,(n)} \, g^{1-n}
$ where \cite{Beisert:2006ib}
\beq
d_{r,s}^{\,(n)}=\frac{\zeta (n) \left((-1)^{r+s}-1\right) \Gamma \left(\frac{1}{2} (n-r+s-1)\right) \Gamma \left(\frac{1}{2}
   (n+r+s-3)\right)}{2(-2\pi)^n \Gamma (n-1) \Gamma \left(\frac{1}{2} (-n-r+s+3)\right) \Gamma \left(\frac{1}{2}
   (-n+r+s+1)\right)}\,.\la{ds}
\eeq
The leading order coefficients at strong coupling are given by
\beq
d^{\,(0)}_{r,s}= \frac{\delta_{s, r-1}-\delta_{s, r+1}}{s\, r}\,\,\,, \qquad d_{r,s}^{\,(1)}=\frac{(-1)^{r+s}-1}{\pi}\frac{1}{r^2-s^2} \,. \la{AFSHL}
\eeq
The simplest way to compute the leading order expression for the dressing factor at strong coupling is to resum (\ref{sumform}) with $c_{r,s}(g)\simeq g\, d^{\,(0)}_{r,s} $. The result is
\be\label{eqeq}
\chi^{(0)}(x,y)=\left(x+1/x-y-1/y\right) \log \left(1-1/xy \right)-1/x+1/y\,.
\ee
The last two terms in (\ref{eqeq}) cancel out when constructing the dressing factor as in (\ref{parame}) while the first term yields
\beq
\sigma(u,v) \simeq \frac{1-\frac{1}{\xm \yp}}{1-\frac{1}{\xp
   \ym}}  \left(\frac{1-\frac{1}{\xm \ym}}{1-\frac{1}{\xm \yp}}\frac{1-\frac{1}{\xp
   \yp}}{1-\frac{1}{\xp \ym}}\right)^{i (v-u)} \la{AFS} \,.
\eeq
This is the so called AFS dressing factor. It was engineered by Arutyunov, Frolov, and Staudacher \cite{Arutyunov:2004vx} to match the strong coupling Bethe Ansatz equations with the KMMZ integral equations \cite{Kazakov:2004qf} describing classical string solutions. Historically, this work was the first solid indication that $\sigma^2(u,v)\neq 1$.

To compute the subleading term at strong coupling, and also the next-to-subleading etc, it is convenient to use the DHM representation (\ref{DHM}) and
\beq
\frac{1}{i}\log \frac{\Gamma(1+i x)}{\Gamma(1-ix)} = -x \log(x/e)^2 - \frac{\pi}{2}\, \sign(x)+\sum_{n>0,\,\, odd}^{\infty} \frac{2 \zeta(-n)}{n}\frac{1}{x^n} \la{gammaexp}
\eeq
which is valid for real $x$. When using this expansion in (\ref{DHM}) we see that the first term yields the AFS dressing factor (\ref{AFS}). The second term, gives us the leading quantum correction, known as the Hernandez-Lopez phase \cite{Hernandez:2006tk}. The sum in (\ref{gammaexp}) yields all other subleading quantum corrections.

Let us now discuss the leading quantum correction. The $\sign$ term in (\ref{gammaexp}) simply constrains the limits of integration in (\ref{DHM}). This leads to 
\beq
\chi^{(1)}(x,y)=\frac{1}{4\pi} \int\limits_{-1}^{1} \frac{dz'}{y-z'} \int\limits_{1/z'}^{z'} \frac{dz}{x-z}-\( x\leftrightarrow y\) \,,
\eeq
which can be directly computed in terms of dilogarithms. In this way we obtain the resummation of (\ref{expansion}) using $d_{r,s}^{\,(1)}$,  performed in \cite{Arutyunov:2006iu}, see also \cite{Beisert:2006ib}. The Hernandez-Lopez phase is a 1-loop effect in the world-sheet strong coupling expansion and can be derived using only (quasi) classical considerations \cite{Frolov:2002av}.
This was done in \cite{Beisert:2005cw,Hernandez:2006tk} using particularly simple circular string solutions \cite{Beisert:2005cw}, checked to be consistent with more complicated solutions in \cite{Freyhult:2006vr}, and derived in full generality in \cite{Gromov:2007cd} using the algebraic curve method \cite{Gromov:2007aq}.

It is fun to notice that the leading and subleading strong coupling terms in the expansion of $\chi(x,y)$ are by far the hardest to compute. All other subleading corrections come from using the last sum in (\ref{gammaexp}) in (\ref{DHM}). They lead to rational integrands in $z$ and $z'$ so that the integrals can be trivially computed by residues.

Another curious feature of the AdS/CFT dressing factor is the following: the strong and weak coupling coefficients (\ref{BEScsg}) and (\ref{ds}) are related by \cite{Beisert:2006ez}
$c_{r,s}^{\,(n)}=d_{r,s}^{\,(-2n-r-s+1)} $.
%Some simple functions do exhibit such remarkable duality between its strong and weak coupling coefficients. It is interesting to observe this behavior for the AdS/CFT dressing factor. 
This relation is further discussed in \cite{Kotikov:2006ts}.

We end this section with the discussion of some other strong coupling checks of the dressing factor.
Explicit perturbative computations of the full S-matrix $\hat S$ were done up to $2$ loops \cite{Klose:2007rz} in the near-flat space limit \cite{Maldacena:2006rv}.
Finite size corrections around the giant-magnon solutions were performed and probe the dressing factor to all loop orders \cite{chapLuescher}
The strong coupling asymptotic expansion of the cusp anomalous dimension was found analytically at any loop order \cite{Basso:2007wd,Kostov:2008ax}. The two loop coefficient was checked through a direct string computation \cite{Roiban:2007dq}.
The reproduction of the $O(6)$ sigma model \cite{Basso:2008tx} from the BES/FRS equation \cite{Freyhult:2007pz} in the Alday-Maldacena limit \cite{Alday:2007mf}, see also \cite{Fioravanti:2008bh,Volin:2010cq}. This probes the dressing phase at all-loop order.
The match of the generalized scaling function
computed at one \cite{Casteill:2007ct} and two \cite{Gromov:2008en}
 loops with a direct string theory prediction at one \cite{Frolov:2006qe} and two \cite{Giombi:2010fa} loops. This is a very nontrivial check since it amounts to matching a non-trivial functional dependence.
    Last three checks emerged from the exhaustive study of anomalous dimensions for twist operators. For more references and a review of twist operators see \cite{chapTwist}.

\section{Concluding remarks} \la{conclusions}

The dressing factor of the AdS/CFT system is a remarkable object with a very non-trivial dependence on the momenta of the scattered particles and on the 't Hooft coupling. It has been impressively scrutinized with remarkable success. Still, there are some challenges to be addressed.

Perhaps the most obvious one is the lack of an independent derivation of the dressing factor purely from gauge theory, without recurring to AdS/CFT. The most significant advantage of the string language is the existence of the notion of Wick rotation which allows us to argue in favor of the crossing relation \cite{Janik:2006dc,Beisert:2005tm,Arutyunov:2006yd}.
On the gauge side the situation is much worse since there is no known meaning for the cross channel at all.  This lack of interpretation on the gauge side is present also in the computation of the spectrum at finite volume. The finite volume computation is based on the Wick rotation trick of Zamolodchikov \cite{Zamolodchikov:1991et} which was implemented for the AdS/CFT case in \cite{Ambjorn:2005wa,Arutyunov:2007tc}. There is no interpretation of the Wick rotation from the gauge theory side, and therefore there is no derivation of the $Y$-system \cite{Gromov:2009tv,chapTBA} which does not rely on the light-cone world-sheet description.

Another interesting puzzle concerns the involved structure of the dressing factor. It motivates us to search for an underlying simpler system. %In many models such simpler system can be indeed constructed. 
For instance, the $O(4)$ sigma model dressing factor can be interpreted as an effective interaction between spin wave excitations in a very simple antiferromagnet  \cite{Faddeev:1985qu}.
There has been some interesting progress in this direction in the AdS/CFT context \cite{Mann:2005ab}.
%
%: Gromov and Kazakov, following \cite{Mann:2005ab}, consider a bath of many relativistic particles in the $O(4)$ sigma model. If we integrate out the physical momenta of the particles we are left with an effective $S$-matrix for the momenta of their isotopic spin waves. The effective dressing factor obtained in this way is precisely the AFS dressing factor (\ref{AFS})!
%It would be wonderful if such construction could be carried through beyond the classical limit. Indeed, this would provide us with the relativistic description of the AdS/CFT integrable system. This description would be in terms of meromorphic functions on the plane instead of the infinite-genus surface described in the text. There were also other attempts to represent dressing factor as an effective object of simpler integrable systems, see \cite{Rej:2007vm}. Unfortunately none of them was completely successful. 
The difference in signs in the denominators of (\ref{fo4good}) and (\ref{goodf}) might be telling us to look for noncompact spin chains.

\section*{Acknowledgments}
Research at the Perimeter Institute is supported in part by the Government of Canada through NSERC and by the Province of Ontario through MRI. The work of D.V. is supported by the US Department of Energy under contracts DE-FG02-201390ER40577. D.V. would also thank for the hospitality of the Perimeter Institute, where the part of this work was done. This work was partially funded by the research grants PTDC/FIS/099293/2008 and CERN/FP/109306/2009.

\phantomsection

\end{document}